\documentclass[aps,prapplied,reprint,amsmath,amssymb,floatfix]{revtex4-2}
\usepackage[utf8]{inputenc}
\usepackage{amsmath}
\usepackage{graphicx}
\usepackage{esint}

\makeatletter

\providecommand{\tabularnewline}{\\}

\usepackage{xspace}
\usepackage{graphicx}

\makeatother

\begin{document}
\title{Meissner-London susceptibility of superconducting\\ right circular
cylinders in an axial magnetic field}
\author{Ruslan Prozorov}
\email{prozorov@ameslab.gov}

\affiliation{Ames Laboratory, Ames, IA 50011, U.S.A.}
\affiliation{Department of Physics \& Astronomy, Iowa State University, Ames, IA
50011, U.S.A.}
\date{started 3 May 2021; revised: 3 July 2021}
\begin{abstract}
Analysis of magnetic susceptibility of non-ellipsoidal samples is
a long-standing problem in experimental studies of magnetism and superconductivity.
Here the quantitative description of the Meissner-London response
(no Abrikosov vortices) of right circular cylinders in an axial magnetic
field is given. The three-dimensional adaptive finite-element modeling
was used to calculate the total magnetic moment, $m$, in a wide range
of London penetration depth, $\lambda$, to sample size ratios. By
fitting the numerical data, the closed-form universal magnetic susceptibility
is formulated involving only sample dimensions and $\lambda$, thus
providing a recipe for determining the London penetration depth from
the accurate magnetic susceptibility measurements. Detailed protocols
of the experimental data analysis using the developed approach are
given. The results can be readily extended to the most frequently
used cuboid-shaped samples.
\end{abstract}
\maketitle

\section{Introduction}

Magnetic susceptibility measurements are widely used to probe magnetic
and superconducting materials. Real-world samples have finite volume
and come in different shapes and aspect ratios, leading to the distortion
of the magnetic field around them. The importance of this demagnetization
correction was recognized since the inception of the theory of electromagnetic
fields in the second half of the 19$^{\textnormal{th}}$ Century, see
Refs.{[}1-9{]} in Ref.\citep{Chen1991}. It was soon realized that
the simple constant demagnetizing correction is only applicable to
the ellipsoids where the internal magnetic field differs from the
applied but is uniform throughout the volume. In non-ellipsoidal bodies,
the demagnetizing field, $H_{d}$, varies significantly on the surface
and inside the sample on macroscopic scale. In addition, it was obvious
that the demagnetization depends on the intrinsic magnetic susceptibility,
$\chi_{0}$, of the material. To simplify the matter, the majority
of works considered either a fully magnetized state with constant
(saturation) magnetization density, a linear paramagnet where the
local magnetization density, $M\left(r\right)$, could be defined
or a perfect diamagnet with $\chi_{0}=-1$, which also formally applies
to a hypothetical perfect superconductor with the London penetration
depth, $\lambda=0$. The superconductor, however, presents an additional
problem because it is inherently non-local in the electromagnetic
sense. One cannot define the local magnetization, $M\left(r\right)$,
or susceptibility, $dM/dH$ the same way it is done for linear magnetic
materials simply because Meissner screening currents are macroscopic
and flow around the entire sample. Therefore local susceptibility
does not have any particular meaning. However, it is still possible
to define the \emph{effective (integral) demagnetizing factor} via
the total measured magnetic moment, $m$, by writing formally, $\chi=m/\left(VH_{0}\right)$,
where $H_{0}$ is an applied magnetic field far from the sample \citep{Demag2018}.
Moreover, this treatment can be extended to arbitrary London penetration
depth, $\lambda$, but then the complete description of the effective
magnetic response requires another quantity which we call an \emph{effective
sample size}, $R$, which is expressed via the dimensionless \emph{geometric
correction coefficient}, $\eta$.

\subsection{Cylindrical sample geometry}

\begin{figure}[tb]
\centering \includegraphics[width=5cm]{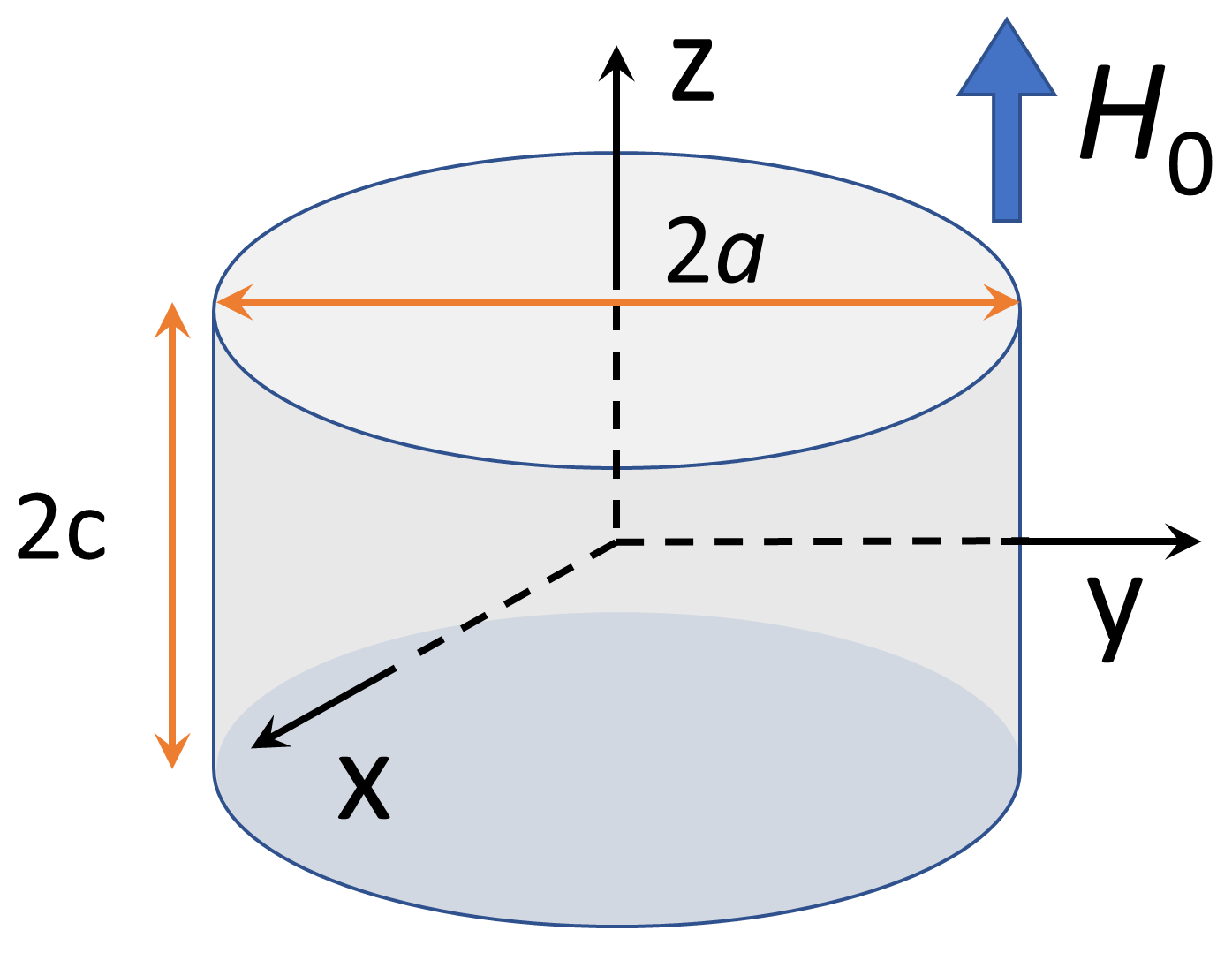} 
\caption{Geometry used in this paper: a right circular cylinder of radius $a$
and thickness $2c$ oriented as shown in a uniform axial magnetic
field, $H_{0}$, produced by external sources.}
\label{fig:fig1}
\end{figure}

We consider geometry, shown in Fig.\ref{fig:fig1}. A right circular
cylinder of radius $a$ and thickness $2c$ (so it has a $2a\times2c$
rectangular cross-section parallel to the magnetic field) is a convenient
for calculations proxy of realistic samples. If $a$ is the actual
sample radius in the plane perpendicular to the field, then the effective
sample radius is defined as $R\left(\nu\right)=\eta\left(\nu\right)a$
where $\nu=c/a$ is the dimensionless aspect ratio with $2c$ being
the sample thickness along the magnetic field. Function $\eta\left(\nu\right)$
together with the effective demagnetizing factor, $N\left(\nu\right)$,
completely determine the total magnetic susceptibility, $\chi$. This
paper gives practical formulas to calculate $\eta\left(\nu\right)$
and $N\left(\nu\right)$ for arbitrary aspect ratio, $\nu$, and shows
how this information can be used for calibration and quantitative
analysis of the experimental DC or AC magnetic susceptibility, for
example, for the estimation of the London penetration depth. The cylindrical
shape is chosen also, because it has a well-defined analytical solution
in the limit of an infinite thickness and eliminates the sharp vertices
in 3D, which may be quite problematic for the numerical work, especially
with finite $\lambda$. Furthermore, cylinders are convenient to handle
in cylindrical coordinates where vector London equation is just a
scalar Helmholtz equation for one component of the vector potential,
$A_{\varphi}$ \citep{Prozorov2000}. 

Of course, real superconducting samples, especially single crystals,
often have platelet shape with two parallel surfaces and polygonal
in-plane cross-section, which varies from rectangles and prisms to
triangles. It turns out, we can approximately map any such shape onto
a cylinder by calculating the radius of an equivalent cylinder based
on the volume penetrated by the magnetic field. For small $\lambda$,
this volume is given by the surface area of crystal faces parallel
to the magnetic field times $\lambda$. Of course, this simplification
neglects top and bottom surfaces perpendicular to the field. However,
for thin samples a renormalized London penetration depth can be introduced
in the same manner. In other words, the penetrated volume is proportional
to the product of sample thickness, penetration depth $\lambda$ and
the perimeter, $P_{S\perp H}$, of the in-plane cross-section, perpendicular
to the field. Therefore, one can map any cross-section shape onto
a cylinder, using $P_{S\perp H}=2\pi a'$, where $a'$ is the radius
of the ``equivalent'' cylinder. For example, often studied cuboidal
sample with a rectangular cross-section, $2a\times2b$, is mapped
onto the equivalent right circular cylinder of radius,
\begin{equation}
a'=\frac{2}{\pi}\left(a+b\right)\label{eq:cuboid_remap}
\end{equation}
and of the same height, $2c$, along the magnetic field, as the original
$2a\times2b\times2c$ cuboid. Therefore, there is a wide applicability
of our results for quantitative calibration of the magnetic susceptibility,
for example to extract the London penetration depth. The step-by-step
procedures are described at the end of the paper.

\subsection{The dimensionality of flux penetration }

As a starting point, let us review some known analytical solutions
of the London equations in different geometries paying attention to
the effective dimensionality of flux penetration. For derivations,
see the mathematically identical problem of the electric field polarization
in conductors of various shapes solved as problems in Chapter VII
of \emph{``Electrodynamics of Continuous Media''}, by Landau and
Lifshitz (vol. 8) \citep{Landau1984}. Everywhere in this paper we
assume isotropic superconductors. 

An apparent measured magnetic susceptibility is defined as: 

\begin{equation}
\chi=\frac{dM}{dH}=\frac{1}{V}\frac{dm}{dH}\label{eq:chi}
\end{equation}
where $M=m/V$ is volume magnetization, $m$ is the total magnetic
moment and $V$ is the sample volume. Note that using volume-normalized
$M$ does not imply spatially-uniform magnetization density. This
magnetic susceptibility $\chi$ is defined only in the integral sense.
For example, two samples of the same volume, but different shapes
will produce two different values of $m$, therefore different $\chi$.
A general discussion of the magnetic moment and its evaluation in
arbitrary shaped samples is given elsewhere \citep{Demag2018}. 

Let us introduce the dimensionality, $D$, of the magnetic field penetration
into the specimen. This is not the geometric dimensionality of the
sample itself (such as 1D wire, 2D film or 3D cube), but rather the
dimensionality of the propagating magnetic flux front advancing into
the sample. Assuming $\lambda\ll R$, the geometric interpretation
of the magnetic susceptibility is that the value of an ideal diamagnet,
$\chi=-1$, is reduced by the ratio of the volume penetrated by the
magnetic field to the total volume of the sample:
\begin{equation}
\chi\left(1-N\right)\approx\frac{\Delta V}{V}-1\label{eq:ChiVol}
\end{equation}
where $N$ is the demagnetizing factor, $V$ is sample volume and
$\Delta V$ is the volume penetrated by the magnetic field. This equation
follows trivially from the exact Eq.\ref{eq:m_total} discussed below.
Let us analyze three representative cases for which the analytic solutions
of the London equation are known.

\emph{One-dimensional} penetration of a magnetic field along the $x-$axis
into a slab of width $2a$ and infinite in $y$ and $z$ directions.
In this case we can think of a volume of a parallelepiped $L\times L\times\left(2a\right)$
with the side surface $L\times L$ parallel to the applied field.
The penetrated volume is:
\[
\Delta V=2\left(aL^{2}-\left(a-\lambda\right)L^{2}\right)=2aL^{2}\frac{\lambda}{a}=V\frac{\lambda}{a}
\]
where a factor of $2$ in front comes from two surfaces located at
$\pm a$. Here the dimensionality of flux penetration is determined
by the term $\left(a-\lambda\right)$ in the first power resulting
in $\chi\sim r$, therefore $D=1$. Equation \ref{eq:ChiVol} gives:
$\chi=r-1$, where $r=\lambda/a$. Indeed, the exact solution is compatible
with this small-$r$ limit \citep{Shoenberg1952},
\begin{equation}
\chi_{slab}\left(r\right)=r\tanh\frac{1}{r}-1\label{eq:slab}
\end{equation}

\emph{Two-dimensional} penetration occurs in an infinite cylinder
of radius $a$ in an axial magnetic field. Following the same logic
as above we can write per length $L$ along the cylinder axis:
\[
 \Delta V=L\pi a^{2}-L\pi\left(a-\lambda\right)^{2}\approx2V\frac{\lambda}{a}
\]
where we only left first order in small parameter, $r=\lambda/a$,
terms. Here the dimensionality of flux penetration is determined by
the $\left(a-\lambda\right)^{2}$ term, so to the first order $\chi\sim2r$,
which implies $D=2$. The exact solution of the London equation for
the infinite cylinder in an axial field is \citep{Landau1984},
\begin{equation}
\chi_{cyl,axial}\left(r\right)=2r\frac{I_{1}\left(\frac{1}{r}\right)}{I_{0}\left(\frac{1}{r}\right)}-1\label{eq:cylinder_axial}
\end{equation}
where $I_{\nu}\left(z\right)$ are the modified Bessel functions of
the first kind. Indeed, it has a factor of 2 in front. It is interesting
to note that the transverse magnetic field penetration into such cylinder
has exactly the same functional form only adding a factor of $2=1/\left(1-N\right)$
from the demagnetizing factor, $N=1/2$ \citep{Landau1984}:

\begin{equation}
\left(1-\frac{1}{2}\right)\chi_{cyl,perp}\left(r\right)=2r\frac{I_{1}\left(\frac{1}{r}\right)}{I_{0}\left(\frac{1}{r}\right)}-1\label{eq:cylinder_perp}
\end{equation}
and the above arguments about $2D$ flux penetration hold in this
case as well.

\emph{Three-dimensional} flux penetration into a superconducting sphere
of radius $a$ is accompanied by the penetrated volume:
\[
\Delta V=\frac{4}{3}\pi a^{3}-\frac{4}{3}\pi\left(a-\lambda\right)^{3}\approx3V\frac{\lambda}{a}
\]
where we dropped all higher order terms in $r$. Here, following the
same logic as above, the effective dimensionality is determined by
the $\left(a-\lambda\right)^{3}$, giving $\chi\sim3r$, which means
$D=3$ in this case. Indeed, the exact solution for a sphere is \citep{Shoenberg1952},

\begin{equation}
\left(1-\frac{1}{3}\right)\chi\left(r\right)=3r\coth\frac{1}{r}-3r{{}^2}-1\label{eq:sphere}
\end{equation}
Similar to the cylinder in transverse field, there is a pre-factor
$\left(1-N\right)$ with $N=1/3$ in this case. In the first two cases
of the infinite slab and cylinder, $N=0$. Clearly, the dimensionality,
$D$, of the magnetic flux penetration and demagnetizing factor are
not rigidly connected. 

\begin{figure}[tb]
\centering \includegraphics[width=8.5cm]{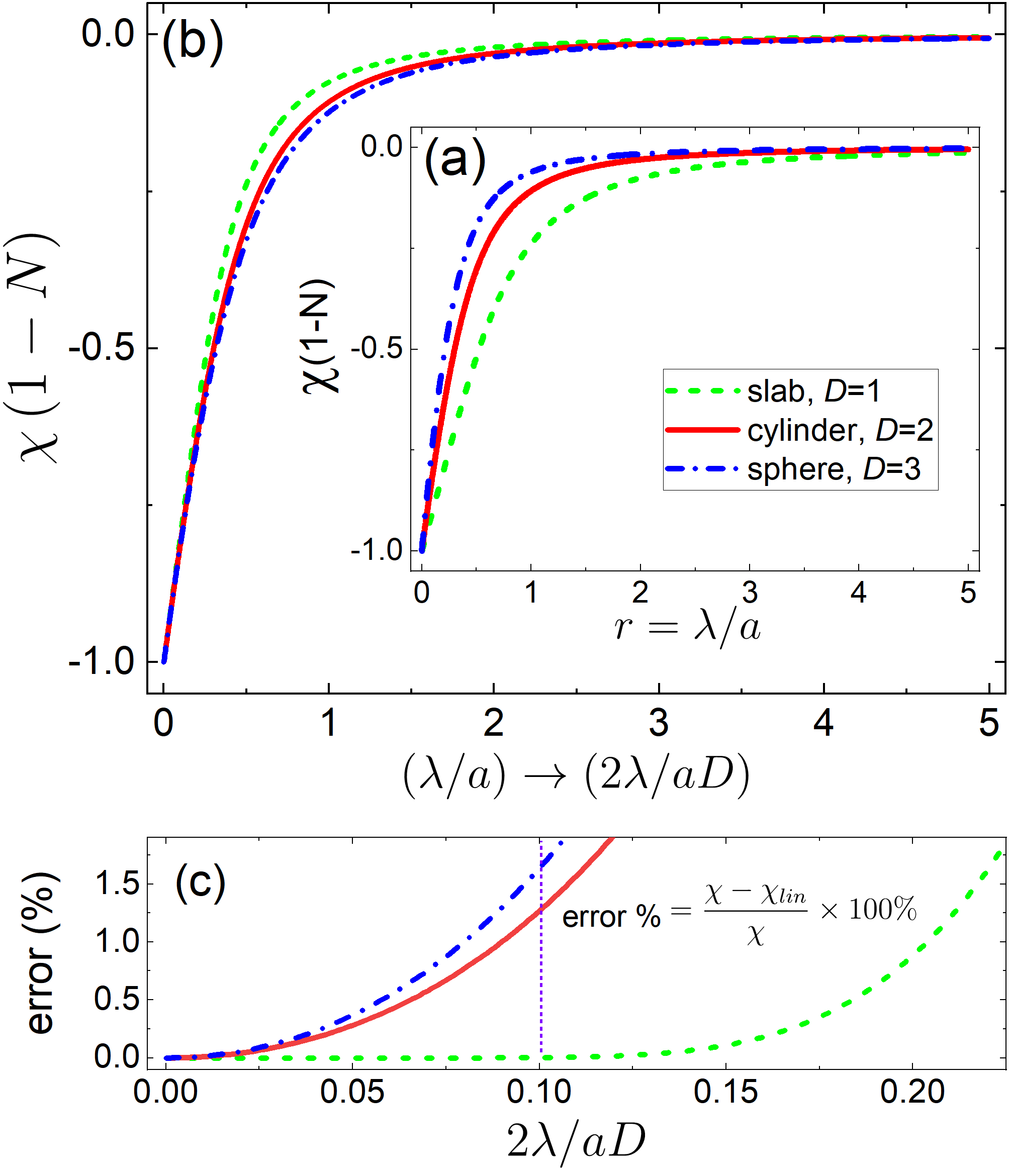} \caption{(a) (inset) Magnetic susceptibility of an infinite slab, an infinite
cylinder (axial and perpendicular - the same curve) and of a sphere
as function of $r=\lambda/R$; (b) the same curves, but in re-scaled
penetration parameter, $r\rightarrow2r/D$; (c) percentage deviation
of the re-scaled curves of the full expressions, Eqs.\ref{eq:slab}-\ref{eq:sphere},
from the linear form, Eq.\ref{eq:linearChi}.}
\label{fig:fig2}
\end{figure}

Figure \ref{fig:fig2}(a) shows these analytic functions, Eqs.\ref{eq:slab}-\ref{eq:sphere},
plotted on one graph revealing quite a substantial apparent difference
at all values of $r$. Obviously, the difference is mostly due to
different $D$, which determines the slope of $\chi\left(r\right)$
at small $r$. Therefore, it is convenient to re-scale the field penetration
parameter $r$ to make the initial slope the same for all curves.
In principle, any suitable function ($\tanh$ or Bessel functions)
and any $D$ can be chosen, it is convenient to choose $D=2$ , because
this is the most general case matching the dimensionality of the magnetic
field itself. It penetrates general sample from two sides and is infinite
in the third direction. Therefore we can write: 
\begin{equation}
r=\frac{\lambda}{a}\rightarrow\frac{2}{D}\frac{\lambda}{a}\label{eq:rescaling}
\end{equation}
which modifies the infinite slab case to $r\rightarrow2r$, leaves
the $2D$ case unaffected, and changes the $3D$ case to $r\rightarrow2r/3$. 

Figure \ref{fig:fig2}(b) shows that the same curves become much closer
to each other when plotted in re-scaled coordinates, Eq.\ref{eq:rescaling},
and coincide (by design) at small values of $r$. In all cases, the
initial susceptibility is apparently linear in $r=\lambda/R$ and
can be written as:

\begin{equation}
\left(1-N\right)\chi_{lin}=D\frac{\lambda}{a}-1\label{eq:linearChi}
\end{equation}

The goal of this paper is to find numerical solutions for the magnetic
susceptibility in finite right circular cylinders with rectangular
cross-section and present it in the form similar to Eq.\ref{eq:linearChi},
as
\begin{equation}
\left(1-N\right)\chi_{lin}=\frac{\lambda}{\eta a}-1\label{eq:linearChiEta}
\end{equation}
where all the effects of finite geometry and deviations from the ideal
shape are absorbed in the effective demagnetizing factor, $N,$ and
the response to the propagating magnetic flux, including the factor
$D=2$, are reflected in the dimensionless function $\eta$, so that
$R=\eta a$ can be viewed as the ``effective'' sample radius.

\subsection{Magnitude of the London penetration depth }

Before we proceed with the calculations, let us estimate at what values
of $r$ the non-linearity caused by the hyperbolic functions become
important. We calculate the percentage deviation as:

\begin{equation}
\textnormal{error}=\frac{\chi-\chi_{lin}}{\chi}\times100\%\label{eq:deviation}
\end{equation}
This quantity is shown in Fig.\ref{fig:fig2}(c) for all three cases.
A quite conservative estimate is that the linear approximation, Eq.\ref{eq:linearChi},
is good within 1\% accuracy for $r\lesssim0.1$. Let us translate
this criterion into the real-world values. Consider a generic temperature-dependent
London penetration depth, $\lambda\left(t\right)$=$\lambda\left(0\right)/\sqrt{1-t^{2}}$,
where typical $\lambda\left(T=0\right)\lesssim300\;\textnormal{nm}$,
and typical $a\approx500\;\mu\textnormal{m}$ for a mm - sized crystal.
Here $t=T/T_{c}$ is the reduced temperature and $T_{c}$ is the superconductor
transition temperature. Then the dimensionless ratio $r=0.1$ is reached
at $t(r=0.1)=\sqrt{1-\left(\lambda\left(0\right)/0.1a\right)^{2}}=0.99998\approx1-\left(2\times10^{-5}\right)$.
This is, indeed, experimentally negligible temperature distance from
$T_{c}$. In fact, the range of realistically important values of
$r$ is much less than $0.1.$ Even a simple estimate of a temperature
at which $\lambda$ increases five fold, $\lambda\left(t_{5}\right)=5\lambda\left(0\right)$
gives $t_{5}=\sqrt{1-\left(1/5\right)^{2}}\approx0.98$, which covers
most of the superconducting domain, while this value corresponds to
only $r=5\times0.3/500=0.003$ where Eq.\ref{eq:linearChi} is definitely
a very good approximation. Note that sometimes used two-fluid phenomenological
function, $\lambda\left(t\right)=\lambda\left(0\right)/\sqrt{1-t^{4}},$gives
estimates even closer to $T_{c}$. Therefore, we conclude that practically
important values of $r$ are small and the linear approximation, Eq.\ref{eq:linearChi}
is well justified in most practical situations. This also eliminates
the problem of temperature-dependent demagnetizing factor (only for
a superconductor!) and allows us to use the demagnetizing factor of
an ideal diamagnetic sample of the same shape and volume as the sample
under study. 

In Section II.D we revisit these estimates, because the
field penetration parameter can be significantly renormalized, especially
in thinner samples. We find that even then linear approximation is
applicable.

\subsection{Numerical solution of the London equations in three dimensions}

\begin{figure*}[t]
\centering \includegraphics[width=17cm]{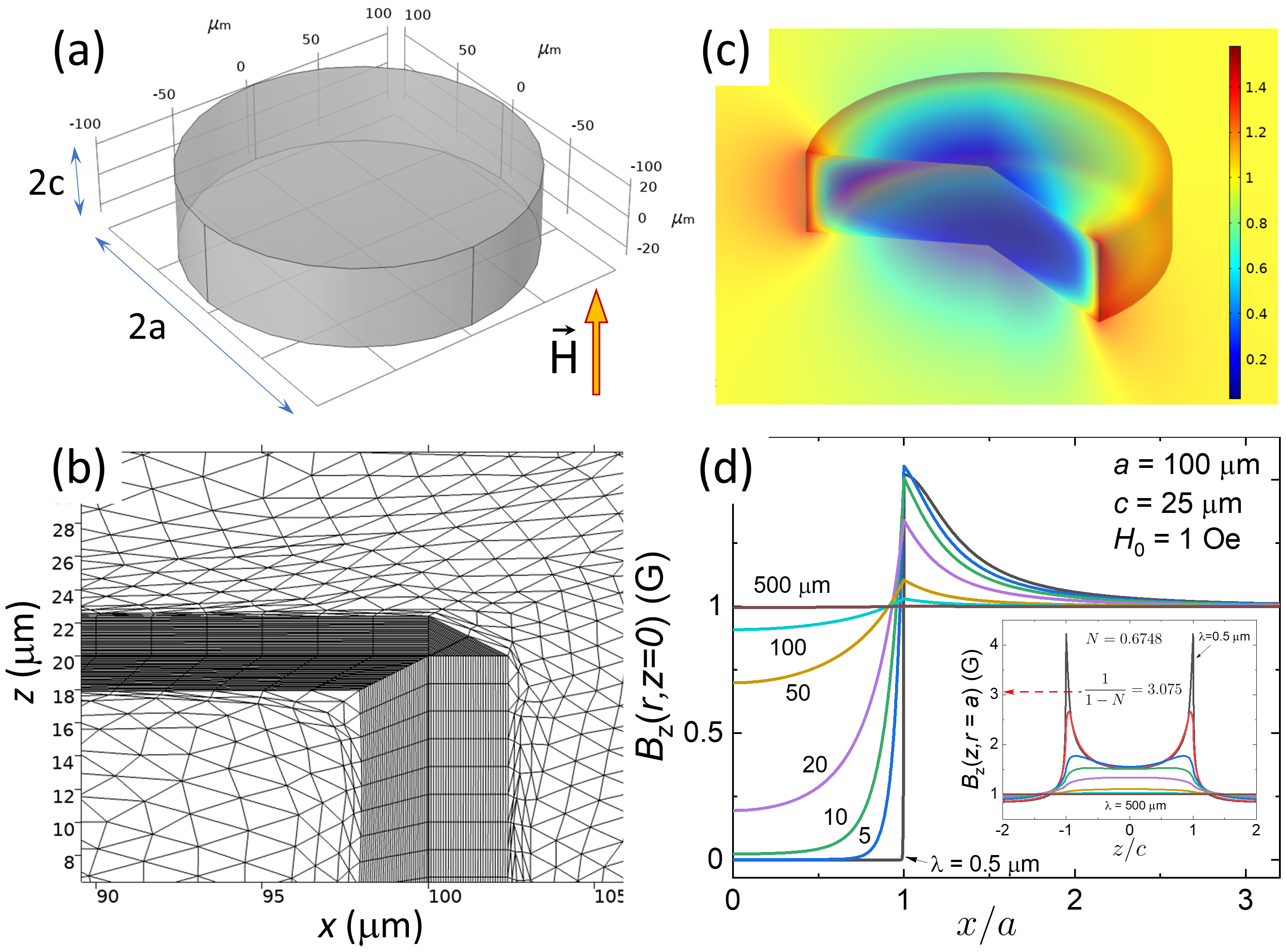} \caption{(a) Geometry defined in COMSOL with designation of dimensions used
in this work. Compare to Fig.\ref{fig:fig1}. (b) Example of the adaptive
mesh near the corner with mesh fine enough to resolve exponential
flux distribution (note the axes scales in $\mu\textnormal{m}$); (c) COMSOL calculation
example of the 3D distribution of the magnetic induction of a superconductor
in an applied magnetic field, $H_{0}=1\;\textnormal{Oe}$, for a large
$\lambda/a=0.2$ to make flux penetration visible. (d) Profiles of
the $z-$component of the magnetic induction calculated for the sample
with $a=100\;\mu\textnormal{m}$ and $c=25\;\mu\textnormal{m}$ at $\lambda=0.5,1,5,10,20,50,100,500\;\mu\textnormal{m}$.
Main frame: $B_{z}(r)$ through the sample center, $z=0$; inset:
$B_{z}(z)$ along the sample side, $r=a$.}
\label{fig:fig3}
\end{figure*}

For numerical computations we use the AC/DC module of COMSOL 5.6 software
\citep{comsol2021} designed to solve the frequency-dependent Maxwell
equations with proper boundary and initial conditions. Detailed description
of the software and its capabilities can be found in the documentation
\citep{comsol2021}. Figure \ref{fig:fig3} (a) shows how the geometry
of the sample, shown in Fig.\ref{fig:fig1}, is implemented in COMSOL.
The Maxwell's Ampère's equation written for a material with pure imaginary
electric conductivity, 
\begin{equation}
\sigma=-\frac{i}{\mu_{0}\omega\lambda^{2}}\label{eq:sigmaSC}
\end{equation}
turns it into the London equation when we set the relative electric
permittivity $\epsilon=0$. The critical step of the numerical analysis
is the construction and optimization of the adaptive mesh that should
be fine enough to resolve the exponential penetration of a magnetic
field into a superconductor at distances smaller than the London penetration
depth, $\lambda$, and it has to be done in a three-dimensional space.
In case of a cylinder the 3D problem of a vector field, $\mathbf{B}=\left(B_{x},B_{y},B_{z}\right)$
can be mapped onto a 2D problem, $\left(B_{r},0,B_{z}\right)$, considering
only one component of the vector potential in cylindrical coordinates
is needed, $\mathbf{A}=\left(0,A\left(r,z\right),0\right)$. Still,
full 3D distribution of the magnetic field is obtained numerically
exploiting the axial symmetry of the problem. Figure \ref{fig:fig3}(b)
shows an example of the small part of the adaptive mesh cross-section,
zooming onto a tiny area near the corner. The actual mesh is even
much denser and this image would be completely black. In the calculations
we used $a=100\;\mu\textnormal{m}$ and calculated $\lambda$ values starting
from $0.1\;\mu\textnormal{m}$ to $500\;\mu\textnormal{m}$ which covers the
realistic range of $\lambda$$\left(T\right)$ in typical superconductors.
We also explored a difficult case of $c<\lambda$$\ll a$ in order
to describe an important thin film limit. 

To check the numerical results, different types of the adaptive mesh
(e.g. tetrahedral or hexahedral) and its refinement strategies was
attempted. Different sample and mesh sizes with and without boundary
layers constructed. The reported finding are robust and independent
on the particular choice of the mesh or algorithms. In all cases,
the extrapolation of the results to the ideal diamagnet limit, $\lambda=0$,
was checked for consistency. In that limit, a much simpler (and therefore
easier to solve) magnetostatic problem of a linear magnetic material
with $\chi=-1$, was calculated. This limit is explored elsewhere
\citep{Demag2018}. 

Figure \ref{fig:fig3}(c) shows an example of a calculated 3D distribution
of the magnetic induction calculated for a particular aspect ratio,
$\nu=c/a=25/100=0.25$, and the London penetration depth, $\lambda=20\;\mu\textnormal{m}$,
in a static applied external magnetic field of $H_{0}=1\;\textnormal{Oe}$.
Figure \ref{fig:fig3}(d) shows the magnetic induction profiles calculated
for the indicated values of $\lambda$. The main frame shows a radial
distribution, $B_{z}\left(r\right)$, from the sample center at $z=0$,
and the inset shows $B_{z}\left(z\right)$ along the sample side,
at $r=a$, for the same set of $\lambda$. Note a very strong variation
of the magnetic field along the surface invalidating any attempt to
use the ``average'' demagnetizing field to map this onto an ellipsoid
where this field is constant, $H=H_{0}\left(1-N\right)$.

\section{Results and discussion}

\subsection{The effective demagnetizing factor, $N$}

The effective demagnetizing factor, $N$, in arbitrary-shaped samples,
including finite cylinders, was discussed in detail elsewhere \citep{Demag2018}.
Here we review the results of these calculations and provide more
detailed formulas for specific cylindrical geometry considered here
to be used when constructing the quantity, $\chi\left(1-N\right)$
from the experimental data. There are several ways to introduce $N$
in non-ellipsoidal bodies. The total magnetic field experienced by
the sample includes the demagnetizing field, $H_{total}=H_{0}+H_{d}$.
In a uniformly magnetized ellipsoid with constant demagnetizing factor
$N$, this field is also uniform, $H_{d}=-NM$, where $M$ is constant
volume magnetization. Inspired by this general picture, different
authors used various numerical, empirical and analytical methods to
evaluate $\mathbf{H}_{d}\left(\mathbf{r}\right)$ and $\mathbf{M\left(r\right)}$,
which are no longer uniform in the non-ellipsoidal samples, see Fig.\ref{fig:fig3}(d). 

In this paper we calculate the total magnetic moment, $m$, in the
direction of the applied magnetic field, which is only straightforward
in the case of free currents of known distribution. However, it is
also possible to calculate $m$ integrating the known distribution
of $B_{z}\left(r\right)$, which is relatively simple numerically,
but its proof in general samples is non-trivial Ref.\citep{Jackson2007,Demag2018}.
The magnetic moment (and an apparent ``integral'' magnetic susceptibility
are given by):

\begin{equation}
\mathbf{m}=\alpha\intop_{V}\left(\frac{\mathbf{B}\left(\mathbf{r}\right)}{\mu_{0}}-\mathbf{H_{0}}\right)d^{3}\mathbf{r}=\chi\mathbf{H_{0}}V\label{eq:m_total}
\end{equation}
where the coefficient $\alpha=3/2$ for the integration over the large
spherical domain that includes the whole sample, whereas $\alpha=1$
is for the integration domain as a large cylinder with its axis parallel
to $\mathbf{H_{0}}$. (Yes, the integral depends on the way it is
evaluated! Mathematically, this is similar to finding an electric
field in a uniformly charged space \citep{Demag2018}). Then, using
the definition:
\begin{equation}
\chi=\frac{\chi_{0}}{1+N\chi_{0}}\label{eq:chi-vs-chi0}
\end{equation}
the effective (integral) demagnetizing factor is defined as, 

\begin{equation}
N=\frac{1}{\chi}-\frac{1}{\chi_{0}}\label{eq:N_general}
\end{equation}
where $-1\leq\chi_{0}\leq\infty$ is the material's intrinsic magnetic
susceptibility (in a sample without demagnetization). In the case
of a perfect superconductor (perfect diamagnet), $\chi_{0}=-1$ and 

\begin{equation}
N=\frac{1}{\chi}+1\label{eq:N_SC}
\end{equation}

It should be noted that historically, two effective demagnetizing
factors were introduced: $N_{m,f}=\left\langle H_{d}\right\rangle _{V,S}/\left\langle M\right\rangle $$_{V,S}$,
were index $V$ is for the volume averaging, and index $S$ is for
the averaging of these quantities in the middle plane perpendicular
to the field. The first kind is called the ``magnetometric'' factor,
$N_{m}$, and the second is called the ``fluxmetric'' (or ``ballistic'')
demagnetizing factor, $N_{f}$, respectively. Yet another approach
is to use the total magnetostatic energy, $NM^{2}/2\mu_{0}$, which
can be calculated numerically \citep{Sato89}. With three possible
orthogonal orientations of the magnetic field, along $x,y,z$, three
principal demagnetizing factors are defined. Detailed calculations
in finite cylinders with rectangular cross-section are given in Refs.\citep{Taylor1960,Chen1991}.
Chen, Brug, and Goldfarb give a comprehensive discussion with many
references to the prior works \citep{Chen1991}. Unfortunately, perhaps
due to the intricacy of the expressions of the spatial distributions,
most works also assume that the sum of the three demagnetizing factors
in three principal directions, $\sum_{i}N_{i}\equiv N_{x}+N_{y}+N_{z}=1$.
However, this is only true for the ellipsoidal shapes where the magnetic
induction (therefore local susceptibility $\chi$) inside the sample
is constant and uniform. This condition does not hold for the generalized
effective demagnetizing factors, as followed from works that did not
use this assumption \citep{Taylor1960,Chen1991,Demag2018}. The actual
variation of the magnetic field in finite samples is highly non-uniform
(see, e.g., Fig.\ref{fig:fig3}(d)), which makes the fluxmetric factor
to be a very poor approach in general. With the mentioned assumption,
$\sum N_{i}=1$, the magnetometric factor is also incorrect. 

As a particular case, let us compare the effective demagnetizing factor
of a perfectly diamagnetic circular cylinder of radius $a$ with the
square cross-section in the $r-z$ plane, $2a\times2a$, in the axial
magnetic field along the $z$ axis. Our numerical result is $N=0.36569$
\citep{Demag2018}, identical value of the best fit, Eq.\ref{eq:N_exact},
and almost the same $N=0.36576$ from what we call a ``practical''
fit, Eq.\ref{eq:N_practical}. This has to be compared with $N=0.36482$
given by Brandt \citep{Brandt2001} and $N=0.3692$ by Taylor \citep{Taylor1960},
apparently showing an excellent agreement between these studies. Furthermore,
for the sum, Taylor finds: $N_{z}+2N_{x,y}=0.3692+2\times0.3669=1.103>1$,
in agreement with our result, $\sum N_{i}=0.36569+2\times0.36673=1.100$.
In a stark contrast, the works that do use the $\sum N_{i}=1$ assumption
underestimate the demagnetizing factor, for example Sato \emph{et
al.} report $N_{z}=0.307054$ \citep{Sato89} and Arrott \emph{et
al.} give $N=0.311577$ \citep{Arrott1979}. The fluxmetric (ballistic)
factors are much lower due to the significant underestimate of the
demagnetizing field evaluated only in the middle plane where it is
the smallest, e.g., Joseph \emph{et al.} report $N=0.2322$ \citep{Joseph1966}.
This if not unique to cylinders. For example, for a cube along three
principal directions, $N_{x}=N_{y}=N_{z}=0.38919$ from our work \citep{Demag2018}
and $N=0.389667$ calculated by Pardo, Chen and Sanchez \citep{Pardo2004}
showing excellent agreement despite the fact that they used quite
different numerical methods. And, indeed, the sum is $\sum N_{i}=3\times0.38919=1.1676>1$,
notably exceeding 1. 

\subsection{Demagnetizing factor of a right circular cylinder in an axial magnetic
field}

\begin{figure}[tb]
\centering \includegraphics[width=8.5cm]{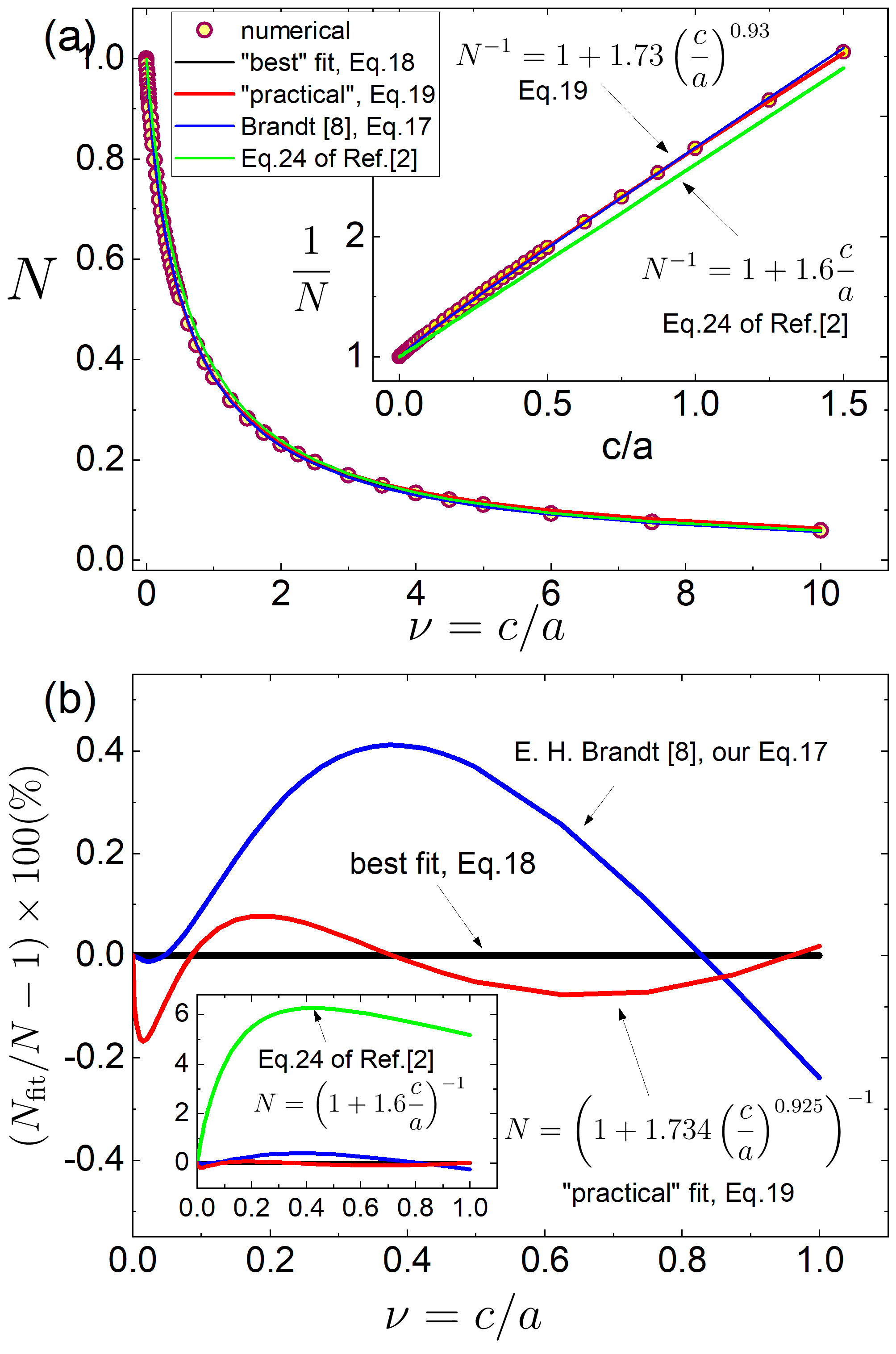} \caption{(a) Effective demagnetizing factor, $N$, as function of the aspect
ratio, $\nu=c/a$. Symbols are the numerical results from our early
work \citep{Demag2018} and solid lines are the fits. The black solid
line labeled ``best fit'' is Eq.\ref{eq:N_exact}, red line ``practical
fit'' is Eq.\ref{eq:N_practical} and blue line is Brand's result,
Eq.\ref{eq:N_Brandt}. For comparison, a simplified curve where $1/N$
is a linear function of $\nu$ is shown by the green line, Eq.24 of
Ref.\citep{Demag2018}. (b): The percent deviation of these fitting
curves from the numerical data. The inset shows a much larger deviation
when linear approximation for $1/N$ is included.}
\label{fig:fig4}
\end{figure}

Unlike the magnetometric factor mentioned above, most often used in
literature, the demagnetizing factor used here involves the actual
value of the applied field, not the average at the sample location
quantity. This uniform magnetic field is set by the external sources
far from the sample. Of many previous works, it turns out that Ernst
H. Brandt has correctly calculated $N$ in this case for arbitrary
aspect ratio \citep{Brandt2001}. His approximate expression for a
cylinder in an axial magnetic field is:

\begin{equation}
\frac{1}{N}=1+\frac{3\pi\nu}{4+2\tanh\left[1.27\nu\ln\left(1+\nu^{-1}\right)\right]}\label{eq:N_Brandt}
\end{equation}
where $\nu=c/a$ is the aspect ratio. Figure \ref{fig:fig4} shows
$N$ evaluated numerically using Eqs.\ref{eq:m_total},\ref{eq:N_SC}
in comparison with approximate expressions. Brandt's Eq.\ref{eq:N_Brandt},
shows an excellent agreement with our numerical results. This, in
turn, lends further support to our calculations considering Brand's
considerable expertise in theoretical superconductivity, in particular
dealing with finite size non-ellipsoidal samples. Brandt has also
noted that $\sum N_{i}\neq1$ for non ellipsoidal shapes.

To find easy to use functional form, we used TableCurve2D software
that fits a given set of data points to dozens of different functions
and ranks them by type and goodness of fit criterion. The best not
too complicated fit of our numerical results for all values of $\nu$
is given by:

\begin{equation}
N=\frac{a+c\nu^{0.5}+e\nu+g\nu^{1.5}+i\nu^{2}}{1+b\nu^{0.5}+d\nu+f\nu^{1.5}+h\nu^{2}}\label{eq:N_exact}
\end{equation}
with the coefficients:\medskip{}

\begin{tabular}{ll}
$a=$ & $0.999999949328650188$\tabularnewline
$b=$ & $3.36894653579048123$\tabularnewline
$c=$ & $3.36942413926991188$\tabularnewline
$d=$ & $8.32140656517217575$\tabularnewline
$e=$ & $5.94222295811372722$\tabularnewline
$f=$ & $7.0090557033581253$\tabularnewline
$g=$ & $-0.363915848221631658$\tabularnewline
$h=$ & $7.5967537756996193$\tabularnewline
$i=$ & $0.034309628991176873$\tabularnewline
\end{tabular}

\medskip{}
Figure \ref{fig:fig4}(b) shows the percentage deviation between the
fit and the numerical results. The black curve labeled ``best fit''
showing a perfect agreement in the whole range from ultra-thin limit
to long cylinders. Equation \ref{eq:N_exact} is the most accurate
choice for numerical work. However, a much simpler and concise (red
curve labeled ``practical'') fit also shows an agreement better
that $0.2\%$ in the whole range,
\begin{equation}
N=\frac{1}{1+1.73401\nu^{0.92505}}\label{eq:N_practical}
\end{equation}

\subsection{Magnetic susceptibility of finite right circular cylinders in an
axial magnetic field}

\begin{figure}[tb]
\centering \includegraphics[width=8.5cm]{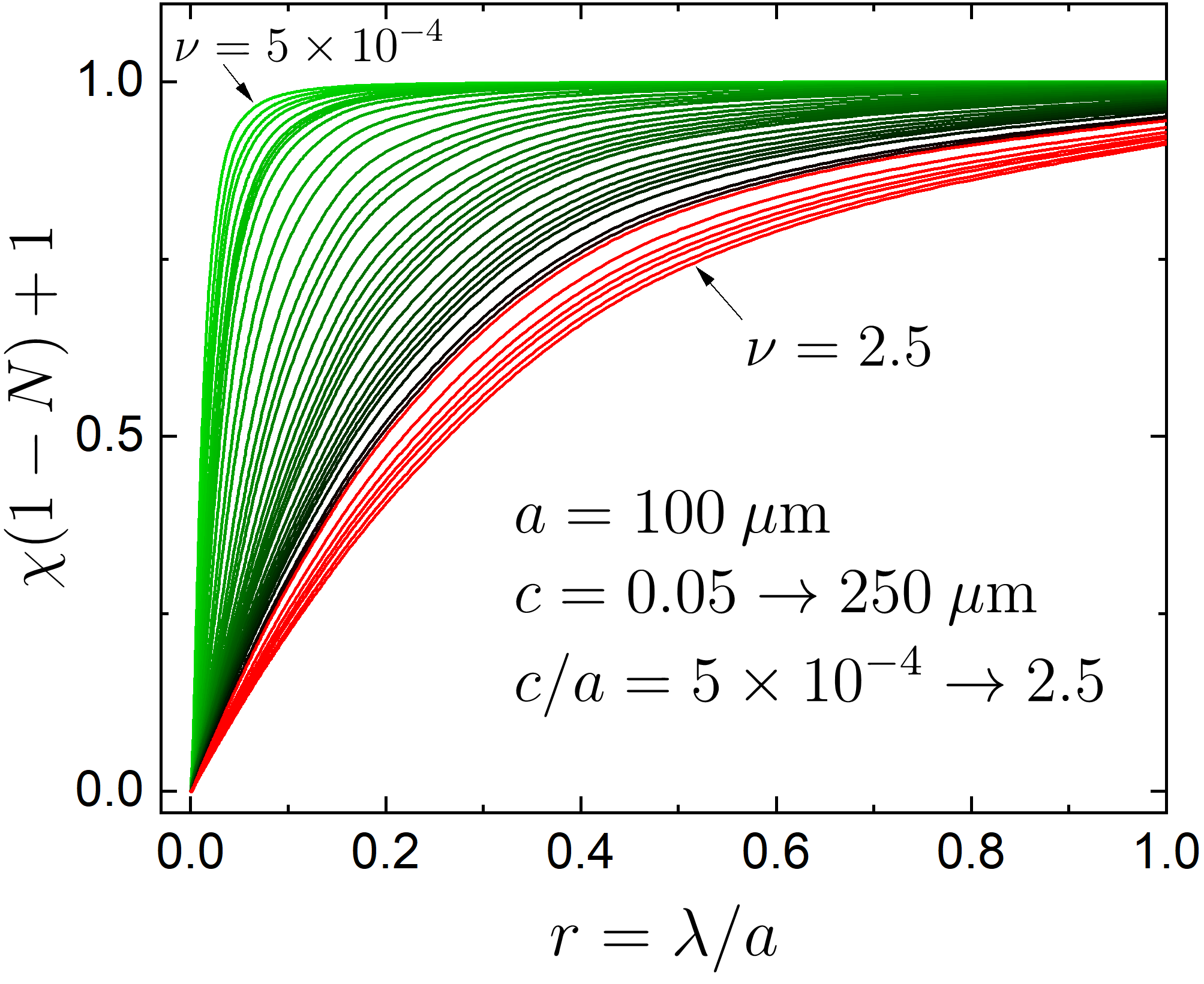} \caption{Calculated magnetic susceptibility of finite cylinders in axial magnetic
field. Plotted is $\chi\left(1-N\right)+1$ versus the ratio $\lambda/a$.
Each curve is calculated for a specific aspect ratio, $\nu=c/a$ raging
from $\nu=5\times10^{-5}$ to $\nu=2.5$, - forty two curves in total. }
\label{fig:fig5}
\end{figure}

The next task is to calculate numerically the apparent magnetic susceptibility
defined by Eq.\ref{eq:chi}. For each value of the ratio, $\nu=c/a$,
the magnetic induction distribution, $\mathbf{B}\left(\mathbf{r}\right)$,
in the sample and space outside is calculated for a wide range of
different values of the London penetration depth (see, for example,
Fig.\ref{fig:fig3}(c) and d)), including the thin film limit of $c<\lambda\left(0\right)\ll a$
as well as long samples up to $c/a=2.5$. Then Eq.\ref{eq:m_total}
was used to obtain the total magnetic moment, $m$, from which the
effective demagnetizing factor was calculated using Eq.\ref{eq:N_SC}.
It is important that we do have a perfect diamagnet limit, $\lambda=0$,
from the solutions for the demagnetizing factor \citep{Demag2018}.
Therefore, we could verify that the calculated curves $\chi\left(r=\lambda/a\right)$,
indeed, extrapolate to the value of $1/\left(1-N\right)$ for $r\rightarrow0$
Therefore, all curves plotted as $\chi\left(1-N\right)$ start from
the same value of $-1$. For convenience of presentation, we start
from zero by plotting $\chi\left(1-N\right)+1$ in Fig.\ref{fig:fig5}.
Each curve shows magnetic susceptibility vs. the ratio of $\lambda/a$
in the interval from $0$ to $2.5$. Each curve is calculated for
a given aspect ratio, $\nu=c/a$, forty two curves total covering
all practically reasonable values from very thin to a very thick limit.

We note that the alternative description would be to include the finite-$\lambda$
dependence in the demagnetizing factor, $N\left(\lambda\right)$,
defined from the general Eq.\ref{eq:N_general}. However, we prefer
to separate the ideal demagnetizing factor, $N$, which depends only
on the sample shape from the characteristic field-penetration parameter,
$\eta\left(\nu\right)$, discussed in the next section. The problem
is that at large values of $\lambda/a$ the equations become non-linear,
instead of Eqs.\ref{eq:slab}-\ref{eq:sphere}. Yet we can keep using
constant $\chi-$independent $N$ and push all the correction to this
parameter $\eta\left(\nu\right)$. Here we introduce the geometric
correction factor, $\eta\left(\nu\right)$, that re-normalizes the
$\lambda/a$ parameter depending on the sample geometry.

\subsection{The geometric correction factor, $\eta\left(\nu\right)$}

\begin{figure}[tb]
\centering \includegraphics[width=8.5cm]{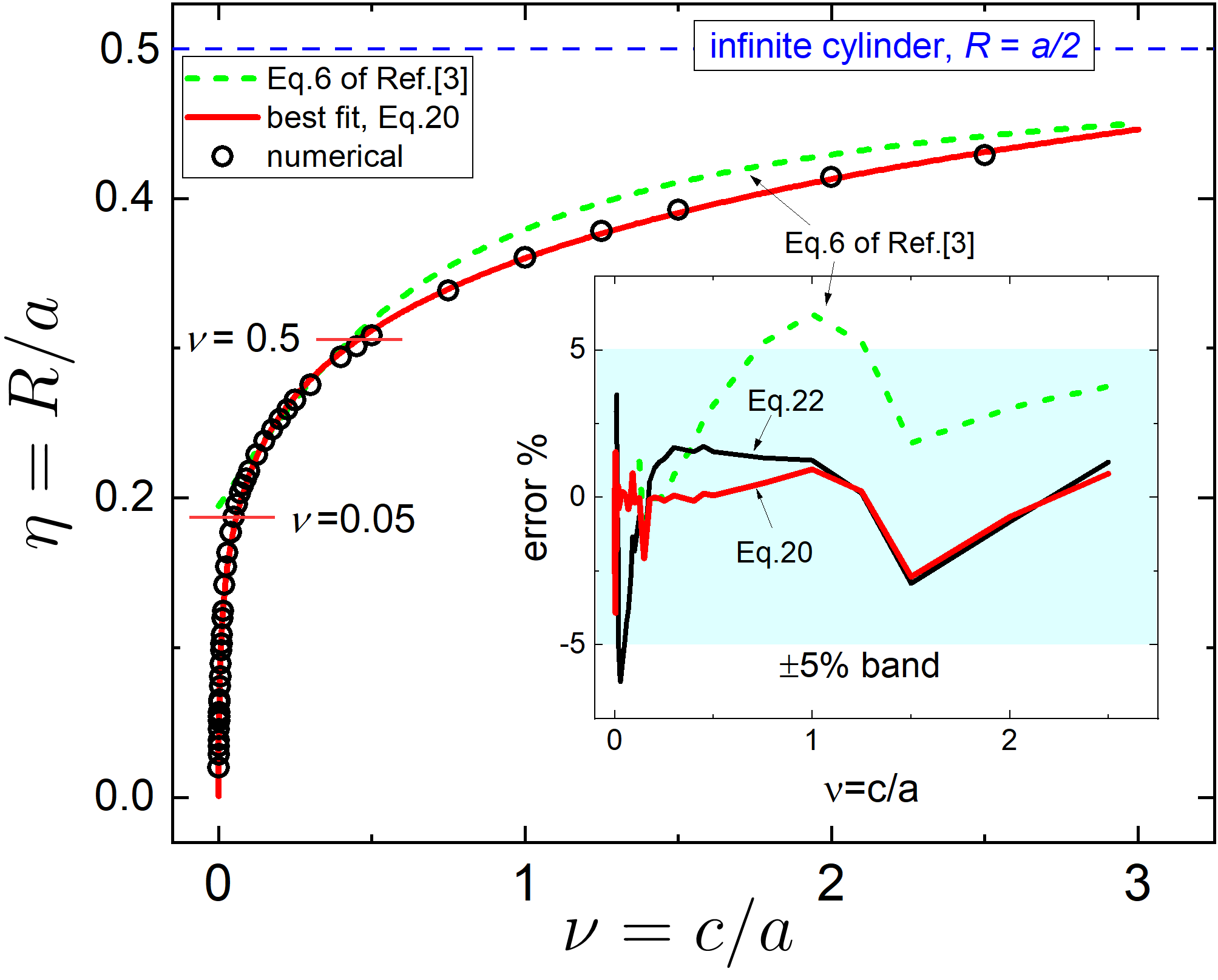} \caption{The geometric correction coefficient, $\eta\left(\nu\right)$ as function
of the cylinder aspect ratio, $\nu=c/a$, calculated by fitting the
initial slope of $\chi\left(\lambda/R\right)$ (see Fig.\ref{fig:fig5})
to Eq.\ref{eq:linearChi} with $D=2$. The solid red line shows our
best fit, Eq.\ref{eq:eta_exact} and the green line shows the plot
of Eq.(6) of Ref\citep{Prozorov2000}. It works not too bad in some
most useful range of aspect ratios, from from approximately $c/a=0.05$
to $0.5$. The thin limit, though is wrong and it is important to
correct that. The simplified fit, Eq.\ref{eq:eta_simple}, is not
shown in the main panel as it is hard to see behind the curves, but
it works very well as can be seen from the inset. The inset shows the relative error between the
fits and the numerical data, $error\;\%=\left(\eta_{fit}/\eta-1\right)\times100\%$.
The red curve corresponds to the best fit, Eq.\ref{eq:eta_exact}.  and black curve to Eq.\ref{eq:eta_simple}.}
\label{fig:fig6}
\end{figure}

As a next step, the initial susceptibility, the slope of each curve
in Fig.\ref{fig:fig5} was fitted to Eq.\ref{eq:linearChi} to obtain
the geometric correction coefficient, $\eta\left(\nu\right)$, that
is used to remap all the curves to one with a common initial slope.
The dimensionality of flux penetration was fixed to $D=2$ to ensure
the correct infinitely-long cylinder limit, $\eta\left(\infty\right)=0.5$.
The red solid curve in Fig.\ref{fig:fig6} corresponds to the best
fit, Eq.\ref{eq:eta_exact}. The symbols are the fitted data points.
The simplified fit, Eq.\ref{eq:eta_simple}, is not shown in the main
panel as it is hard to see behind the curves, but it practically coincides
with the best fit. The discrepancy between the fits and the numerical
results, $\textnormal{error}\;\%=\left(1-\eta_{\textnormal{fit}}/\eta\right)\times100\%$,
is plotted in the inset. For comparison with the previous results,
the green dashed curve shows $\eta\left(\nu\right)$ suggested by
us 21 years ago based on the specific analytical approximation of
the spatial distribution of the magnetic induction on the sample surface
(Eq.(2) of Ref.\citep{Prozorov2000}) inspired by the projection of
an ellipsoid onto a plane and calculating the expelled volume resulting
in Eq.(6) in Ref.\citep{Prozorov2000}. Back then we did not look
at the thinner samples and, as a result, the suggested formula does
not work at all below roughly $\nu<0.05$. Nevertheless, as can be
seen in Fig.\ref{fig:fig6}, the dashed green curve is quite close
to our new calculations approximately in the interval from $c/a=0.05$
to $0.5$. Fortunately, it covers the most practical range of experimental
aspect ratios. For example, for a typical crystal of $1$ mm in diameter,
the thicknesses from $50$ to $500\;\mu\textnormal{m}$ are described
fairly well by Eq.(6) of Ref.\citep{Prozorov2000} and these values
are commonly used in the experiment. Importantly, the Eq.(6) of Ref.\citep{Prozorov2000}
was used extensively for the calibration of the magnetic susceptibility
from sensitive frequency-domain measurements \citep{Prozorov06,prozorov2011}.
In the thin limit, the previous estimate is wrong and it is important
to correct that. Instead of a predicted saturation at $\eta$$\left(0\right)$=0.2,
the correct curves decreases rapidly towards zero. 

The calculated geometric correction factor, $\eta$$\left(\nu\right)$,
can be well represented by the following numerical approximation (solid
red curve in Fig.\ref{fig:fig6}):

\begin{equation}
\eta\left(\nu=\frac{c}{a}\right)=\frac{a+c\nu^{0.5}+e\nu+g\nu^{1.5}+i\nu^{2}}{1+b\nu^{0.5}+d\nu+f\nu^{1.5}+h\nu^{2}+j\nu^{2.5}}\label{eq:eta_exact}
\end{equation}
with the coefficients:\medskip{}

\begin{tabular}{ll}
$a=$ & $-0.0281027472950638618$\tabularnewline
$b=$ & $4.58797296004528776$\tabularnewline
$c=$ & $2.875513314557569$\tabularnewline
$d=$ & $\text{\ensuremath{-137.801329718142642} }$\tabularnewline
$e=$ & $-33.7758403444390059$\tabularnewline
$f=$ & $1249.78954939821191$\tabularnewline
$g=$ & $\text{\ensuremath{188.948638333931676}}$\tabularnewline
$h=$ & $-123.317937748885354$\tabularnewline
$i=$ & $250.500066383141833$\tabularnewline
$j=$ & $138.717299394700242$\tabularnewline
\end{tabular}

\medskip{}
Please note, that while this equation provides a very good approximation
of the calculated numerical values, it is only applicable in the calculated
range of $\nu\leq3$, which in practice covers all aspect ratios of
experimental importance. At $\nu=3,$ Eq.\ref{eq:eta_simple} gives,
$\eta\left(3\right)=0.43963288651773$ while the terminal value at
$\nu\rightarrow\infty$ should be $\eta\left(\infty\right)=0.5$,
marked by the blue dashed line in Fig.\ref{fig:fig6}. For a smooth
extrapolation that is valid for all values of $v$ we use Eq.\ref{eq:eta_exact}
for $\nu\leq3$ and for $\nu>3$: 

\begin{equation}
\eta\left(\nu\right)=0.5-\frac{0.18110134044681}{\nu}\label{eq:eta-inf}
\end{equation}

We also found the simpler fitting function that works quite well for
$\nu\leq3$ :

\begin{equation}
\eta\left(\nu\right)=(a+b\ln\nu)^{2}\label{eq:eta_simple}
\end{equation}
where the coefficients are:\medskip{}

\begin{tabular}{ll}
$a=$ & $0.6013634241567849$\tabularnewline
$b=$ & $0.05987611403976185$\tabularnewline
\end{tabular}

\medskip{}

\begin{figure}[tb]
\centering \includegraphics[width=8.5cm]{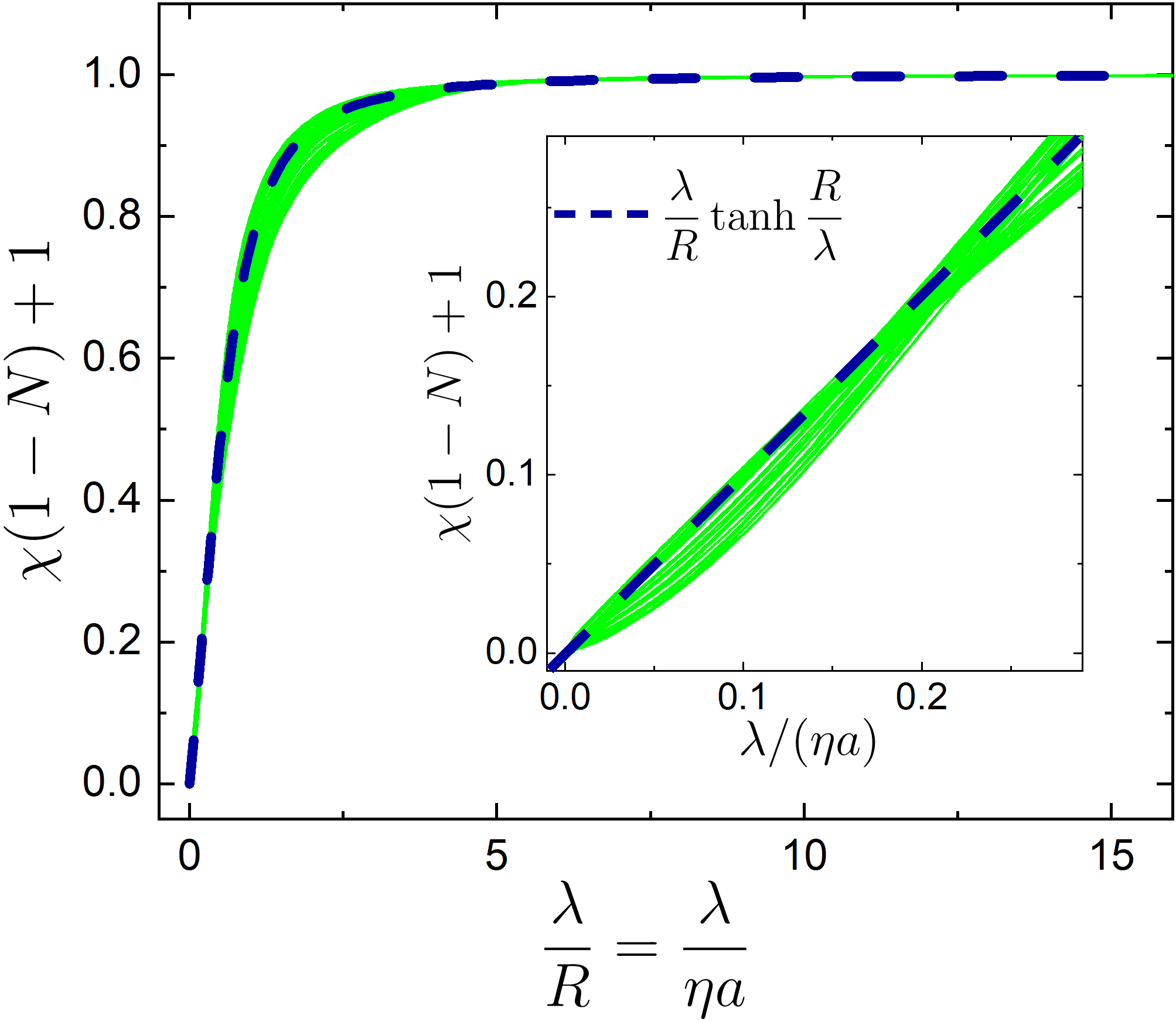} \caption{Scaled magnetic susceptibility, $\chi\left(1-N\right)+1$, plotted
versus the scaled magnetic field penetration ratio, $\lambda/R=\lambda/\left(\eta a\right)$.
Main frame is the full range and the inset shows the linear part.
There are forty two curves plotted together showing the generally
good trend towards the scaled behavior. }
\label{fig:fig7}
\end{figure}

We can now use the calculated geometric correction factor, $\eta$$\left(\nu\right)$,
and the demagnetizing factor, $N\left(\nu\right)$, to plot the scaled
susceptibility, $\chi\left(1-N\right)$ versus the scaled field penetration
ratio, $\lambda/R=\lambda/\left(\eta a\right)$. This is shown in
Fig.\ref{fig:fig7}. Note the large expansion of this parameter range
in thin samples while the calculations were done only up to $\lambda/a=2.5$.
The dashed dark blue line shows analytical $\lambda/R\tanh\left(R/\lambda\right)$
function. Clearly, the scaled curves are clustered around it at all
values and approach a single slope for small $\nu$. This region is
zoomed at in the inset. Of course, the deviations are expected for
the thinner samples due to both, technical and geometrical reasons.
Technically it is difficult to build a good fine adaptive mesh in
very narrow domains and, physically, the response changes when thickness
becomes comparable to the London penetration depth. In fact, we can
now revisit the regime of the linear behavior, Eq.\ref{eq:linearChi}.
We found above that linear approximation is applicable for up to $r\leq0.1$.
Now we need to apply it to the re-normalized quantity, $\lambda/\left(\eta a\right)\leq0.1$.
This places the upper limit on the penetration depth, $\lambda\leq0.1\eta a$,
which gives a hard cutoff when the smallest value, $\lambda\left(0\right)$,
is used. With the same parameters as before, $\eta=\lambda/0.1a=0.3/\left(0.1\times500\right)=0.006$,
which is achieved at $\nu=1\times10^{4}$ or $c=50$ nm for a 1 mm
sized sample. This is a very thin film limit and clearly, our effective
approach used here is no longer applicable. Perhaps, this is a good
way to define an ultra-thin limit. Still, it also means that our approximation
is valid in a wide domain of practical aspect ratios.

\begin{figure}[tb]
\centering \includegraphics[width=8.5cm]{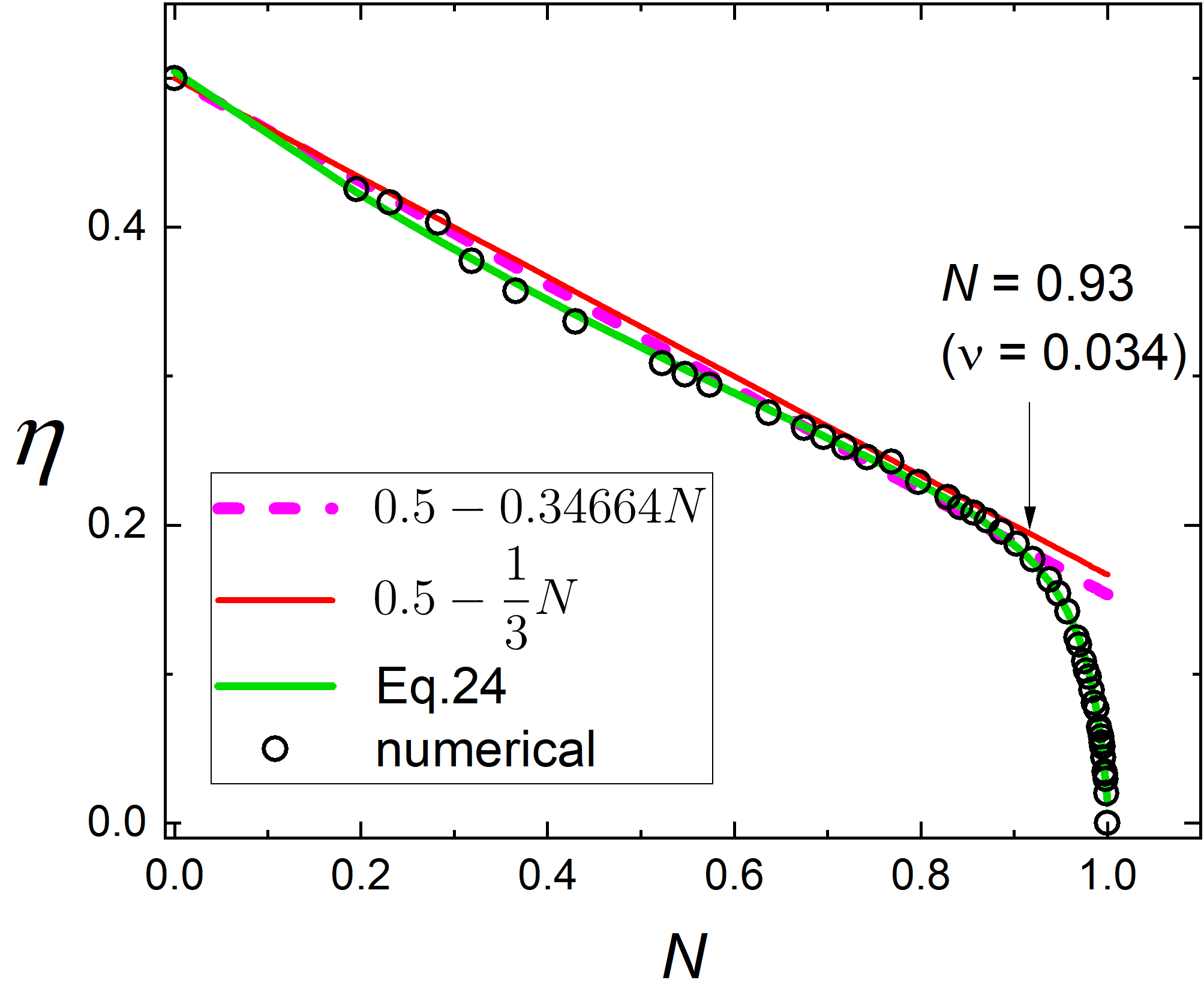} \caption{Geometric correction coefficient, $\eta$, as function of the demagnetizing
factor, $N$. Symbols are the numerical results, dashed magenta line
is the best linear fit and the red line is simple approximation, $\eta=\frac{1}{2}-\frac{N}{3}$
in the interval $0\protect\leq N\protect\leq0.93$. The green curve
is the reasonable fit valid at all values of $N$, see Eq.\ref{eq:eta-vs-N}.}
\label{fig:fig8}
\end{figure}

There is another way to analyze our results along the lines discussed
at the end of Section II.B. Note that both, demagnetizing factor,
$0\leq N\leq1$, and the geometric correction coefficient, $0\leq\eta\leq0.5$,
are limited to the indicated intervals while the aspect ratio, $v=c/a\geq0$
does not have an upper bound. Therefore, we can examine the function,
$\eta\left(N\right)$ and see whether it has some simple representations
and trends. Figure \ref{fig:fig8} shows the numerical results along
with three approximations. Apparently, $\eta\left(N\right)$ is linear
in the interval, $0\leq N\leq0.93$. The best linear fit is $\eta\left(N\right)=0.5-0.34664N$
and, obviously, the simplest possible interpolation is:
\begin{equation}
\eta\left(N\right)=\frac{1}{2}-\frac{N}{3},\;\textnormal{for\;}0\leq N\leq0.93\label{eq:eta_N_linear}
\end{equation}
The whole curve is described reasonably well by the function:

\begin{equation}
\eta\left(N\right)=\frac{a+cN+eN^{2}}{1+bN+dN^{2}}\label{eq:eta-vs-N}
\end{equation}
with the coefficients:\medskip{}

\begin{tabular}{ll}
$a=$ & $0.504378594750316222$\tabularnewline
$b=$ & $-0.517078267930805991$\tabularnewline
$c=$ & $-0.709566812021391927$\tabularnewline
$d=$ & $-0.446415444537349392$\tabularnewline
$e=$ & $0.205676653892788686$\tabularnewline
\end{tabular}

\medskip{}
It will be interesting to perform similar calculations for different
geometries and see whether Eq.\ref{eq:eta-vs-N} will show some universality
or not. 

\section{examples of practical applications}

The obtained numerical results and analytical approximations can be
used for general calibration of precision magnetic susceptibility
measurements. Here we give two examples of practical applications.

\subsection{Finding $N$ and $\eta$ without direct measurements of sample dimensions}

The important utility of Eq.\ref{eq:eta-vs-N} is that it gives the
simple connection between $N$ and $\eta$, bypassing the need to
figure out the exact dimensions of the crystal. Considering realistic
shapes, calculating volume, $V$, from sample dimensions is no trivial
task. On the other hand $V$ can be often estimated from the weight
far more accurately than from the direct size measurements. (Material
density is usually known at least from the parameters of the unit
cell in crystals). Then, at the lowest temperature (as close to $\chi_{0}=-1$
as possible), one can use SQUID (or any other sensitive) magnetometer
to measure the total magnetic moment, $m$, in an external magnetic
field, $H_{0}$. Assuming perfect diamagnetism (which is quite accurate
already below $0.98T_{c}$ as described in the introduction), the
demagnetizing factor can be estimated from Eq.\ref{eq:m_total} and
Eq.\ref{eq:chi-vs-chi0},
\begin{equation}
N=\frac{VH_{0}}{m}+1\label{eq:N-from-chi}
\end{equation}
Now using Eq.\ref{eq:eta-vs-N} the geometric correction factor $\eta$
can be calculated. These two numbers, $N$ and $\eta$ are unique
for each sample and they allow for a complete quantitative description
of the magnetic properties in the Meissner-London state. Indeed, one
can check the result by evaluating $N$ and $\eta$ using sample dimensions
and compare the result. Whether Eq.\ref{eq:eta-vs-N} has some universality
or scaling in different shapes remains to be seen. It requires conducting
similar calculations for other shapes, for example cuboids. If we
compare demagnetizing factor for a cube, $N_{cube}\left(\nu=1\right)=0.38919$
with the demagnetizing factor for a cylinder with square cross-section,
$N_{cylinder}\left(\nu=1\right)=0.36569$ they are both greater than
that of a spheroid (sphere in this case) $N_{sphere}=1/3$, but the
limiting case of all three for infinite $v$ is the same, $N_{\infty}=0$.
For the cylinder with the square cross-section, $\eta_{cylinder}\left(\nu=1\right)=0.35714$
and for the sphere, by the definition of the re-scaling discussed
in Section I.A, $\eta_{sphere}\left(\nu=1\right)=2/3=0.6(6)$, and
therefore, it is more likely that the curves are different for each
geometry. However, it is still possible that these curves for different
shapes will only differ by a simple factor, because the general form
of Fig.\ref{fig:fig8} will be similar due to the bounds of $N$ and
$\eta$.

\subsection{Measurements of the London penetration depth}

Suppose some quantity, $f,$ proportional to the magnetic susceptibility
is measured. This can be the frequency of a tunnel-diode resonator
or of a resonant microwave cavity, or the AC voltage of a sensitive
susceptometer or, the DC voltage of a SQUID circuit with flux transformer.
When a sample is inserted into the coil that provides an excitation
magnetic field, $H_{0},$ (AC or DC depending on the method) small
enough to stay in the Meissner state (no Abrikosov vortices), the
measured quantity changes by some amount, $\varDelta f$. The maximum
possible change, $\text{\ensuremath{\varDelta f_{0}}}$, is induced
when an ideal diamagnet $(\lambda=0)$ of the sample size and shape
is inserted. Therefore, the actual magnetic susceptibility is
\begin{equation}
\left(1-N\right)\chi=-\frac{\varDelta f}{\varDelta f_{0}}=\frac{\lambda}{R}-1\label{eq:calib_lambda}
\end{equation}
where the effective sample dimension, $R=\eta\left(\nu\right)a$.
The quantity $\varDelta f_{0}$ can be measured directly by pulling
the sample out of the coil at the base temperature. (Yes, one of our
cryostats is equipped with a mechanical pulley of a spring-loaded
sample holder, thus we can measure $\varDelta f_{0}$ directly and
compare with the calculations). Alternatively, it can be calculated
using the demagnetization factor described above. For example, in
the case of a resonant circuit frequency shift,

\begin{equation}
\varDelta f_{0}=\frac{f_{0}V}{2V_{c}\left(1-N\right)}\label{eq:df0}
\end{equation}
where $V$ is sample volume and $V_{c}$ is volume of the coil. See
Refs.\citep{Prozorov2000,Prozorov2000a} for detailed discussion and
derivation of Eqs.\ref{eq:calib_lambda} and \ref{eq:df0}. Now one
can use Eq.\ref{eq:N_exact} or \ref{eq:N_practical} to calculate
$\varDelta f_{0}$. The next step is to extract the London penetration
depth, $\lambda$, which is the final goal of this calibration procedure.
In reality, the uncertainty of macroscopic dimensions compared to
$\lambda$ is so large that it is only possible to precisely measure
the relative changes with respect to the base temperature value, roughly
speaking $\varDelta\lambda=\lambda\left(T\right)-\lambda\left(0\right)$,
see detailed discussion in Ref.\citep{Prozorov2000a}. In other words,
Eq.\ref{eq:calib_lambda} becomes:
\begin{equation}
\delta f\left(T\right)\equiv\varDelta f\left(T\right)-\varDelta f_{0}=\varDelta f_{0}\frac{\varDelta\lambda\left(T\right)}{R}\label{eq:calib_delf}
\end{equation}
The measured quantity, $\delta\left(T\right)$ now gives the change
in the London penetration depth,

\begin{equation}
\varDelta\lambda\left(T\right)=\frac{R}{\varDelta f_{0}}\delta f\left(T\right)=G\delta f\left(T\right)\label{eq:calib_Dlambda}
\end{equation}
where $R=\eta\left(\nu\right)a$ is the effective sample dimension
calculated from Eq.\ref{eq:eta_exact} (or approximately from Eq.\ref{eq:eta_simple}).
For example, in the case of a tunnel-diode resonator, $f_{0}\sim10\;\textnormal{MHz},$$\varDelta f_{0}\sim5\;\textnormal{kHz}$
for typical sub-mm sized crystals with $R\sim100\;\mu\textnormal{m}$.
This gives the calibration constant $G=R/\varDelta f_{0}\sim20\;\textnormal{nm}/\textnormal{Hz}$.
The smallest frequency changes practically detected are about $\delta f\sim0.02\;\textnormal{Hz}$
(giving $2$ parts per $10^{9}$ sensitivity of this technique!) resulting
in resolving $\varDelta\lambda\left(T\right)\sim0.4\;\textnormal{\AA}$
which is small enough to study the anisotropy of the superconducting
gap \citep{Prozorov06,prozorov2011}. The full London penetration
depth requires another measurement of the absolute value at least
at one temperature \citep{Prozorov2000a}.

\section{Conclusions}

In conclusion, a systematic quantitative description of the magnetic
susceptibility of superconducting right circular cylinders of arbitrary
aspect ratio, $\nu=c/a$, in an axial magnetic field is outlined for
finite London penetration depth. It includes two independently computed
quantities. (1) The demagnetizing factor, $N\left(\nu\right)$, of
the ideal diamagnetic sample of the same shape and volume as the sample
of interest and (2) the geometric correction factor, $0\leq\eta\left(\nu\right)\leq0.5$,
which takes into account the penetration of the magnetic field from
all sides, including top and bottom surfaces, thus effectively reducing
the relevant effective dimension, $R=\eta\left(\nu\right)a$, where
$a$ is the radius of the cylinder. A simple connection between $N$
and $\eta$ valid for not too thin samples, $\nu\gtrsim0.03$, is
given by: $\eta\left(N\right)=0.5-N/3$. It may be helpful for a quick
estimate of the geometric factor determining the effective flux penetration.
With these quantities, the universal magnetic susceptibility is given
approximately by:
\begin{equation}
\left(1-N\right)\chi=\frac{\lambda}{R}\tanh\frac{R}{\lambda}-1\label{eq:chi_universal.}
\end{equation}

The developed approach can be used for the quantitative analysis of
the magnetic susceptibility data to extract the London penetration
depth, $\lambda$, and these results can be extended to the cuboid-shaped
samples for the approximate estimates. 
\begin{acknowledgments}
I thank Vladimir Kogan for ongoing insightful discussions and John
Kirtley for sharing his COMSOL code when I was learning this software.
I also thank all members of my group, especially Makariy Tanatar and
Kyuil Cho for various input, discussions and support. This work was
supported by the U.S. Department of Energy (DOE), Office of Science,
Basic Energy Sciences, Materials Science and Engineering Division.
Ames Laboratory is operated for the U.S. DOE by Iowa State University
under contract DE-AC02-07CH11358.
\end{acknowledgments}

\bibliographystyle{apsrev4-2}
%

\end{document}